# Structure of Exposure Modeling of Systems to Climate Change Related Hazards

*An instantiation with Boolean Exposure Modeling applied to physical assets facing extreme temperature scenarios.*

Matthieu Dutel[a,b], Adam Abdin[a], Didier Soto[b],

Anne Barros[c]

[a]*Laboratoire Génie Industriel, CentraleSupélec, Université Paris-Saclay, Gif-sur-Yvette, France*
[b]*Resallience by Sixense Engineering, Nanterre, France*
[c]*Chair on Risk and Resilience of Complex Systems, Laboratoire Génie Industriel, CentraleSupélec, Université Paris-Saclay, Gif-sur-Yvette, France*

---

Physical infrastructures designed for stable climate conditions support a significant portion of daily human life activities. However, climate change reshapes the distributions of extreme climate intensities. As a result, physical infrastructures that may face extreme events exceeding their design thresholds risk damage and require adaptation. A preliminary step to adaptation is to identify exposed infrastructures to climate hazards. This paper aims to structure the exposure modeling to climate change hazards by providing an overall methodology. This paper instantiates a framework using the Boolean Exposure Model to assess how physical systems may be exposed to climate hazards. The framework combines a Boolean System Model for infrastructures and a Boolean Hazard Model informed by extreme value statistics. An illustrative case study quantifies the exposure of road bridges in French administrative units to extreme temperatures. Results indicate increasing exposure across future climate scenarios and time periods, with spatial non-uniformity. This framework supports multi-level analysis and integrates non-stationary climate extremes, providing a structured exposure model to facilitate the adaptation of physical systems to climate change.

# 1. Introduction

**List of Symbols**

**Flowcharts**

|  |  |
|---|---|
| Circle | Starting or ending node of the process. |
| Rectangle | Function, Model, Treatment. |
| Cylinder | Data |
| Rounded rectangle | Physical Reality |
| Diamond | Choice, Selection |

## 1.1. Context and Motivation

For a non-negligible portion of the world's population in industrial areas, daily life activities are dependent on physical infrastructures. To travel from Lyon, France, to Paris, France, for a one-week business trip, various physical infrastructures are utilized, including wastewater treatment plants, roads, buildings, electric lines, data centers, communication networks, railroads, and road bridges. Using these physical infrastructures seems simple because they all have a high level of performance. However, climate change can affect this level of performance. At the macro scale, climate change implies an increase in the mean world temperature (IPCC, 2021). At the local scale, this change affects the distribution of climate intensities and extremes, but not uniformly across all locations. As physical infrastructures are mainly designed considering stable climate extreme conditions, the changes to climate intensity extremes risk damaging these infrastructures and lowering their performance levels. Exposure modeling and vulnerability modeling are crucial components in estimating potential physical damage, as they characterize the interaction between infrastructure and the intensity of the hazard. In this paper, we focus exclusively on exposure modeling. An exposure modeling ideal goal would be to have an exposure scenario of the given infrastructure to a specific hazard with a structured and common method. This scenario would be for each physical infrastructure, for each climate hazard, for each time period and radiative pathway scenario, and for each location.

Considering these facts and the problem raised, we must consider the following questions:

- Is it possible to design a structured exposure model that provides information about the exceedance or non-exceedance of the intensity threshold that may cause damage to a given physical infrastructure type?
- As an illustrative example, is it possible to quantify the exposure scenarios of road bridges to the exceedance of their design limit by extreme temperatures?

## 1.2. Framework and Related Works

This study develops a systematic framework to quantify the exposure of physical systems to climate change-related hazards. The approach integrates climate projections, extreme value statistics, a Boolean Hazard Model, a Boolean System Model, and a Boolean Exposure Model. The approach also enables the use of projected data, the exploration of multiple emission scenarios, parameter fitting on time slices to capture the non-stationarity of the climate, and its application to existing infrastructure assets. While the illustrative use case considers Eurocode-based thresholds, the framework itself is general and can be applied to different infrastructure types, hazards, and design standards.
Previous works have primarily focused on either exposure theory and mapping or hazard analysis within design standards, but rarely on both in an integrated, asset-level, and climate-dependent framework that can be adapted to different infrastructure types and various climate-related hazards.

In the literature on the design standard, Rianna et al. (2023), proposed procedures to update Eurocode thermal loads using extreme value statistics, yet without applying climate projections or real infrastructure data. Our approach extends this work by explicitly linking extreme statistics to projected climate data and by focusing on actual physical infrastructures, while remaining general beyond Eurocode-specific applications. Athanasopoulou et al. (2020) studied future temperature extremes with a RCP8.5 pathway using a multi-model ensemble but did not quantify exposure to real infrastructures. We developed a hazard model and linked projected extremes to system exposure using Boolean Exposure Modeling. Markova et al. (2024) fitted statistical distributions to observed extremes at some locations in the Czech Republic without applying them to projections or an ensemble of existing assets. Our study incorporates the statistical step at the national scale and includes physical asset exposure within an end-to-end exposure pipeline.
In the literature on exposure modeling, Van Westen and Greiving (2017) define exposure as the "spatial overlay of hazard footprints and elements at risk locations". The same concept is adopted by Doan et al. (2023), Papilloud et al. (2020), and Koks et al. (2019). We adopt this definition, conceptualizing exposure as the overlay (or intersection) of the system of interest and hazard at a given location, which we formalize as:

$$E = H \cap S$$

Where $E$ is the exposure of the system to a specific hazard. Where $H$ is the spatial distribution of the hazard. Where $S$ is the spatial distribution of the system. The hazard $H$ is dependent on a Hazard Intensity Threshold (HIT), which itself depends on the system $S$ and on the climate intensity data.

We then implement the models through a Boolean formulation. This enables quantitative assessment of system exposure to climate-related hazards and direct integration with probabilistic or reliability analyses.
We build the current paper based on Dutel et al. (2025), who introduced the Boolean Exposure Model for administrative units. We extend the exposure framework in which BEM takes place by integrating extreme value statistics to capture heavy-tail behavior and by applying it directly to a bridge database. Palu & Mahmoud (2019) used a 5-year climate block; in contrast, we use 30-year periods and integrate hazard intensity thresholds for exposure assessment.
This integrated approach bridges the methodological and practical gaps between hazard modeling, which considers design limits, and exposure modeling, which combines climate-driven hazard generation, extreme statistics, and spatially explicit exposure computation for an ensemble of physical systems. It provides a reproducible, climate-informed framework that is adaptable to various climate hazards, infrastructure types, and territories, as well as design standards, offering exposure scenarios for adaptation studies. This framework is supported by Figure 1. We provide in Figure 1 the main steps of the methodological pipeline we tackle in this paper. It includes: Step 0.A. the Climate Model Selection; Step 0.B. the System Data Selection; Step 1. The Extreme Value Statistics; Step 2. The Selection of the Hazard Intensity Threshold; Step 3. The Hazard Modeling; Step 4. The System Modeling; Step 5. The Exposure Modeling; Step 6. The Analysis of the Quantified Results.

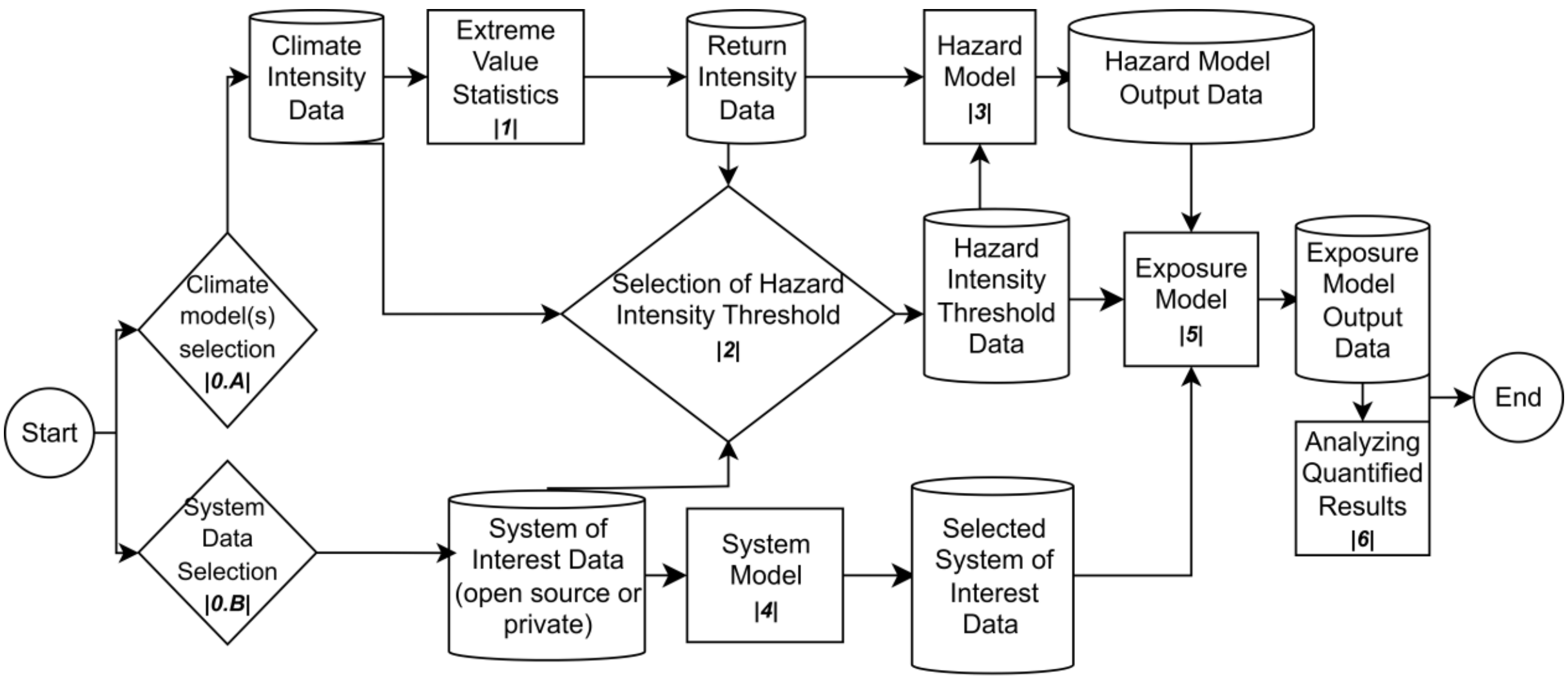

Figure 1. Flowchart of the Exposure Framework methodological pipeline.

1.3. Contribution

This study formalizes a structured and reproducible exposure modeling framework that combines climate projections, extreme value statistics, and physical infrastructures. By applying Boolean models to hazard, physical systems, and exposure, the framework enables an ensemble of asset exposure assessments under evolving extreme climate conditions.

1.4. Paper Structure

The paper is organized as follows: Section 1 introduces the exposure of the system to climate change hazards and situates it within the existing literature. Section 2 presents the model methodology and makes a theoretical and specific model instantiation, as well as hosting the modeling choices. Section 3 presents the practical instantiation of the methodology, thanks to an illustrative use case that traverses different administrative units containing physical assets to a specific ensemble of physical assets: road bridges of SNCF-Réseau facing extreme temperatures. Section 4 presents the results of the models and their instantiation in the illustrative case study. Section 5 concludes the study, discussing what we have accomplished and what could be improved.

# 2. Model Methodology

To address the challenges outlined in Section 1, quantifying infrastructure exposure to climate-related hazards under evolving conditions, we propose a structured modeling framework. This framework integrates climate projections, extreme value statistics, and Boolean models for system, hazard, and exposure. This structured modeling framework enables the assessment of whether physical infrastructures could be exposed to hazard intensities that exceed their design thresholds. This section also outlines the methodological pipeline employed to instantiate this framework, from climate model selection to exposure quantification, thereby enabling reproducible and scalable analysis across various infrastructure types, climate hazards, and territories.

### 2.0.A. Climate model(s) selection

Climate model selection is the initial step. For computational efficiency, this study employs a single modeling framework, though the pipeline is compatible with any model or ensemble. Selection serves to (1) identify the appropriate framework and (2) choose the relevant intensity variable (e.g., temperature, precipitation, wind). While this study focuses on one variable, the methodology can be extended to multiple intensities by iterating on different intensity variables.

### 2.0.B. System(s) & system data selection

System definition is critical and must be implemented in accordance with the study scope, data availability, and resource constraints. Systems may be defined at the macro scale (e.g., assets within French territory) or with precise location and subsystem details. This study uses open-source data as detailed in Section 3.0.B.

### 2.1. Climate data, Return Intensity and Extremum Value Statistics (EVS): Modeling

The extreme value statistics part is summarized in Table.2.1 and illustrated in Figure.2.1.

| Sub Step | Description |
|---|---|
| A. Climate Data | Use of projected gridded climate data; method generalizable to point-based weather station data and gridded historical data |
| B. Block Maxima | Extracts maximum intensity per time block (e.g., month, year). This is required for fitting the Gumbel distribution. |
| C. Dataset split | Data are divided into blocks of equal duration (with a minimum duration of 30 years) for both historical and future periods. |
| D. Extreme Value Model | Fits Gumbel distribution ($\xi = 0$) with location ($\mu$) and scale ( $\sigma$) parameters, and the right-skewed tail allows estimation of the probability of rare, high intensity events. |
| E. Extreme Value Statistics: Non-Stationary Parameter | Non-stationary parameters are central to this study, as we aim to quantify system exposure under climate change. In a stationary climate, such adjustments would be unnecessary. Here, the Gumbel distribution is fitted to each time period and scenario, allowing the hazard model outputs to reflect evolving extreme intensities over time. |
| F. Select Return Intensity | All return intensities can be computed. Since we are in an exposure context, we will select a return intensity relevant for our use case, i.e., for a given type of infrastructure or norms related to a specific infrastructure sector. |
| G. Compute Return Intensities | For a given return period $T$, the return level $i_T$ is the intensity exceeded once every $T$ years. Using Gumbel, this is defined as: $$i_T = \mu - \sigma \times \ln\left(- ln\left(1 - \frac{1}{T}\right)\right) \quad (1)$$ Where $\mu$ and $\sigma$ are the Gumbel location and scale parameters. |
| H. Return Intensity Data | Output is a return intensity $i_T$ level for a given period $T$, the corresponding probability $P_T = \frac{1}{T}$. This will be our input for the rest of the model. |

Table 2.1. Steps to obtain the return intensity.

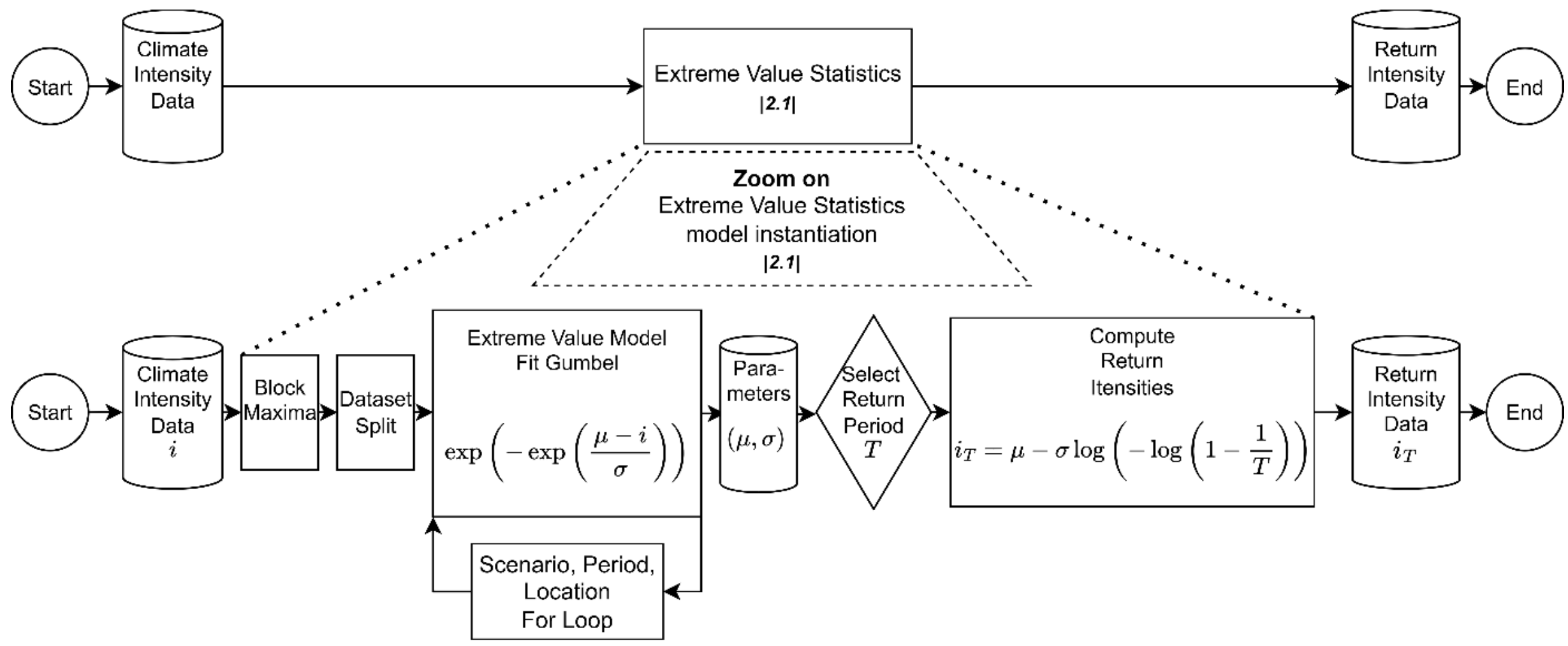

Fig.2.1. Flowchart of the extreme value statistics modeling framework to obtain return intensity for given period(s).

### 2.2. Selection of Hazard Intensity Threshold

As illustrated in Figure 1, the selection of the Hazard Intensity Threshold depends on three key inputs: (i) climate intensity data, (ii) return intensity data, and (iii) system of interest data. The threshold must be expressed in the same physical units as the climate variable (e.g., °C), aligned with the selected return period, and adapted to the system's characteristics. It is therefore a Hazard Intensity Threshold, both climate intensity data-specific and system-specific.

### 2.3. Hazard Modeling: Boolean Hazard Model with Probability

The hazard model takes as input the return intensity data and the hazard intensity threshold data, and the corresponding spatial coordinates (longitude $x$, latitude $y$). The Boolean Hazard Model with Probability (BHMP), introduced in Dutel et al. (2025), we adapt as:

$$H(i_T, x, y) = BHMP_{i_{HIT}, P_{i_{HIT}}}(i_T, x, y) = \begin{cases} 1, & if\ i_T > i_{HIT} + k_{\circ C} \\ 0, & otherwise \end{cases} \quad (2)$$

Where $i_T\ (x, y)$ is the return intensity at location $(x, y)$, $i_{HIT}$ is the Hazard Intensity Threshold at location $(x, y)$, $k_{\circ C}$ is the optional constant (e.g., in °C) for multiclass thresholding. $P_{i_{HIT}}$ is the probability associated with the return period used to define the hazard intensity threshold.

### 2.4. System Modeling: Boolean System Model

The Boolean System Model (BSM) introduced in the work of Dutel et al. (2025), is defined as:

$$S(d_s, x, y) = BSM(d_s, x, y) = \begin{cases} 1, & if\ presence\ of\ a\ system\ of\ interest\ at\ (x, y) \\ 0, & otherwise \end{cases} \quad (3)$$

Where $d_s$ is the data about the system of interest, $BSM$ be the Boolean System Model, $x$ the longitude, $y$ the latitude.

The Boolean System Model can be applied at various levels of physical systems. For example, it may be applied to: a country, regions within a country, or lower administrative units. BSM can also be used for ensembles of physical infrastructure, subsystems within infrastructure, and components of those systems. In the current study we will apply BSM to an ensemble of physical infrastructures.

### 2.5. Exposure Modeling: Boolean Exposure Model

The initial formulation of the Boolean Exposure Model was introduced in Dutel et al. (2025) and is defined as:

$$BEM(i_T, d_s, x, y) = BHMP_{i_{HIT}, P_{i_{HIT}}}(i_T, x, y) \times BSM(d_s, x, y) \quad (5)$$

Where $BEM$ is the Boolean Exposure Model, $BHMP$ the Boolean Hazard Model with Probability, $BSM$ the Boolean System Model, $i_T$ the return intensity data, $d_s$ the system data, $(x, y)$ the spatial coordinates (longitude, latitude).

An equivalent expression, when considering $BEM, BHMP$ and $BSM$ as spatial maps over all locations, is:

$$BEM = BHMP \cap BSM \quad (6)$$

### 2.6. Asset Management Exposure Analysis: Method

To quantify exposure at aggregated levels, we define the exposure rate of a low-level system ($LL$) within a high-level system ($HL$) as:

$$rBEMout_{LL,HL} = \frac{\sum_{i=1}^{N_{LL,HL}} BEMout_i}{N_{LL,HL}} \quad (7)$$

Where : $N_{LL,HL}$ is the total number of low-level systems in a given high-level system, $BEMout_i$ is the Boolean output (0 or 1) of the Boolean Exposure Model for system $i$. For example, the exposure rate of 'communes' within a country is:

$$rBEMout_{Commune,Country} = \frac{\sum_{i=1}^{N_{Commune,Country}} BEMout_i}{N_{Commune,Country}} \quad (8)$$

This aggregated indicator summarizes exposure across different system levels, enabling comparative analysis between systems.

Through this modeling pipeline, we provide a structured approach to address the problem raised in Section 1: can we quantify exposure scenarios of physical infrastructures facing climate extremes that exceed their design limits? The framework is designed to be general and adaptable to various infrastructure types, climate hazards, and design standards or even other types of thresholds. In Section 3, we focus on a specific use case: road bridges exposed to extreme temperatures, illustrating how the framework could be applied in a real-world context.

# 3. Illustrative Case Study: Setup and Results

We demonstrate the practical use of the framework introduced in Section 1 and instantiated in Section 2. The case study focuses on road bridges exposed to extreme temperatures in metropolitan France. The analysis is conducted at the asset level using geolocated infrastructure data, and results are able to be aggregated across administrative units for multi-level interpretation. This case study integrates climate projections, extreme value statistics, and Eurocode-based hazard intensity thresholds to assess exposure, offering a practical instantiation of the modeling pipeline.

### 3.0.A. Climate model(s) selection: illustrative case study setup

To assess exposure to extreme temperature hazards in metropolitan France, we use bias-corrected daily maximum temperatures from the DRIAS-2020 climate service. The data, TasMaxAdjust, is the output of a modeling chain that combines CNRM-CM5 r1 (a global circulation model), ALADIN6.3 v2 (a regional climate model), and ADAMONT France (bias correction and spatial disaggregation). For further technical details, refer to Dutel et al. (2025).

### 3.0.B. System(s) data selection: illustrative case study setup

The Boolean system model uses the longitude and latitude of road bridges. We use an open-source dataset from SNCF Réseau "Liste des ponts-route", which includes bridges where roads cross railway tracks. This dataset covers metropolitan France, except Corsica. Administrative boundaries are sourced from IGN Admin Express. These geolocated bridges serve as the system of interest in our exposure modeling framework, allowing us to quantify which of these assets may face temperature extremes beyond their design thresholds.

### 3.1. Extreme Value Statistic: illustrative case study setup

| Sub Step | Description |
| --- | --- |
| A. Climate Data | Daily maximum temperature, TasMaxAdjust, 8km SAFRAN grid within France's administrative boundaries. |
| B. Block Maxima | Yearly Maxima: for each year and location, the highest temperature is extracted. |
| C. Dataset split | Temperature data is divided into 30-year intervals with a 24-year stride. |
| D. Extreme Value Model | Gumbel distribution is fitted for each location, each 30-year time period, and each scenario. |
| E. Extreme Value Statistics: Non-Stationary Parameter | Gumbel parameters from 30-year blocks, locations, and scenarios capture climate non-stationarity and non-uniform spatial changes in extreme intensities, enabling the calculation of return intensities. |
| F. Select Return Intensity | A 50-year return period is chosen, aligning with Eurocode design standards for infrastructure, such as bridges. |
| G. Compute Return Intensities | Using Gumbel parameters, the 50-year return temperature (probability 0.02) is computed for each location and scenario. |
| H. Return Intensity Data | The output is a 50-year return temperature for each location, scenario, and time period, which is used as input for hazard modeling. |

Table 3.1. Steps to obtain the return intensity in the use case.

### 3.2. Selection of Hazard Intensity Threshold: illustrative case study setup

The Hazard Intensity Threshold (HIT) considers climate data intensity and the physical system under study, as Dutel et al. (2025) explained. The threshold is chosen accordingly to Clause 6.1.3.2 (1) of NF EN 1991-1-5/NA, a French standard based on Eurocode. It provides principles and rules for calculating temperature actions and their effects on structures, including bridges. This standard links extreme temperature data to physical assets: if the standard is respected and no hazard (i.e., no exceedance of the standard) occurs, no damage to assets is expected. However, temperatures exceeding the standard introduce an uncertainty zone, where damage may occur, since such extremes are not considered in the asset's design. While some asset managers may use stricter thresholds, we rely on the Eurocode standard due to its established structural engineering basis and lack of asset-specific data. In this study, the standard acts as a baseline exposure map, linking French departmental territories and their physical assets to static extreme temperatures at the departmental level.

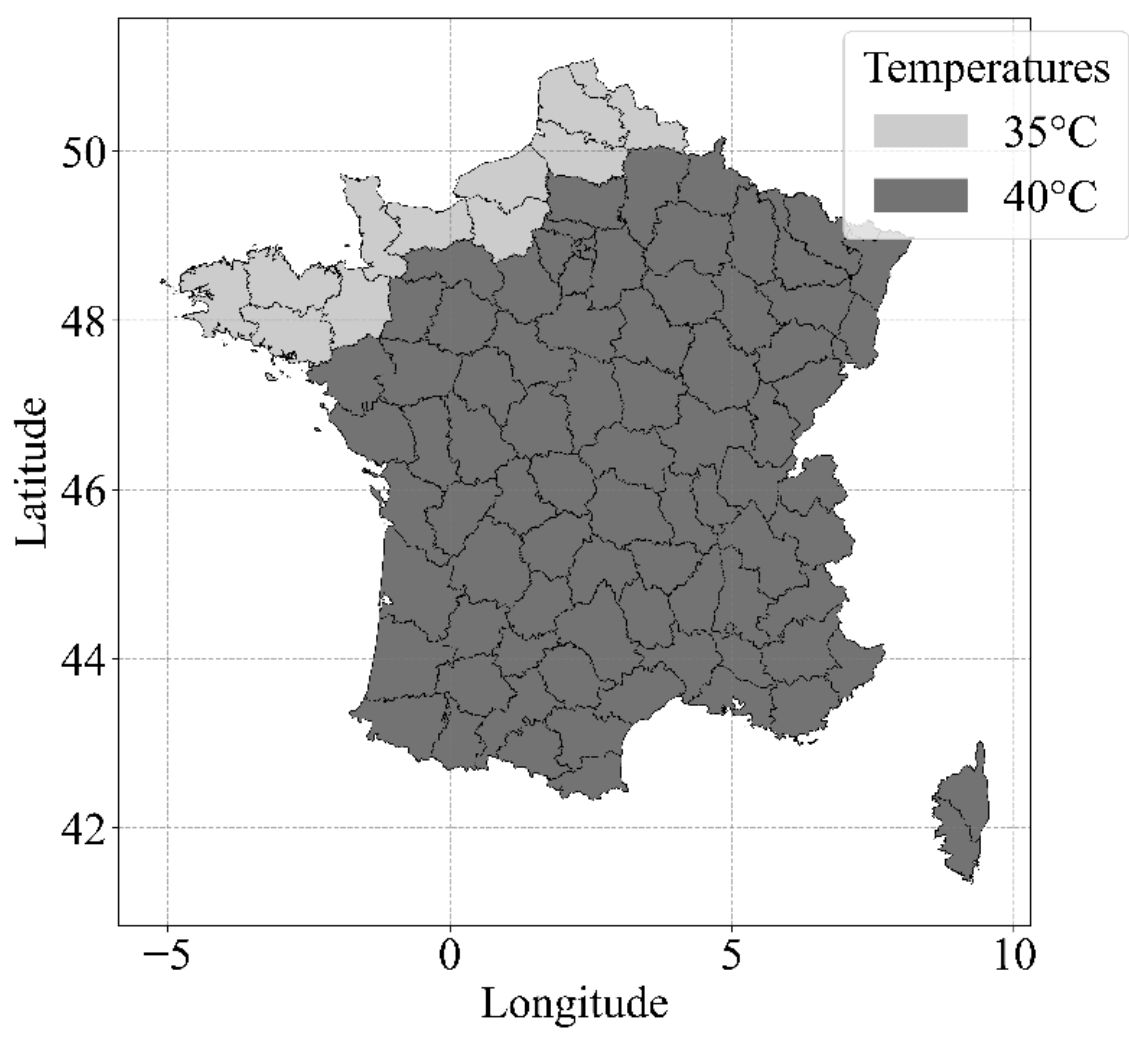

**Fig. 3.2.2.** Maximum air temperature under shelter, by French department from Clause 6.1.3.2 (1) of NF EN 1991-1-5/NA. Figure from Dutel et al. (2025).

### 3.3. Boolean Hazard Model with Probability: illustrative case study setup

To instantiate the Boolean Hazard Model with Probability, we combine two inputs: the spatialized 50-year return temperature intensities ($i_{T50}$) derived from extreme value statistics (see Section 3.1), and the Hazard Intensity Thresholds (HIT) defined at the departmental level (see Section 3.2).
For each scenario, period and location on the SAFRAN grid, the model evaluates whether the projected $i_{T50}$ exceeds the corresponding HIT. If so, a hazard is recorded; otherwise, the location is considered without hazard. This binary classification is repeated across eleven scenario-period combinations, resulting in eleven hazard maps, each covering 8595 points. To support asset prioritization and reflect varying degrees of exceedance, we extend the BHMP to multiclass outputs by applying the following thresholds: HIT + 2.5 °C, HIT + 5 °C, HIT + 7.5 °C, and HIT + 10 °C. These outputs will be linked to the system of interest in the exposure model (Section 3.5) to identify where infrastructure may face climate extremes beyond its design limits.

### 3.4. Boolean System Model: illustrative case study setup

The Boolean System Model is instantiated across multiple system levels, from national to asset scale. As shown in Figure 3.4.1, we consider five levels: country, region, 'département','commune', and the ensemble of SNCF Réseau road bridges. BSM encodes for each level the presence or absence of the system of interest at each location. These inputs can be applied to entire system levels or specific subparts, enabling flexible exposure analysis. This multi-level structure supports comparisons across administrative units and asset ensembles. As demonstrated in Appendix A.1, the spatial distribution of road bridges is non-uniform across administrative units. This non-uniformity is a key consideration when quantifying exposure rates and interpreting results. This BSM instantiation provides the system input needed for the exposure model to operate. Through this exposure method, we can identify, across different territorial levels, which infrastructures may be exposed to climate extremes that exceed their design thresholds.

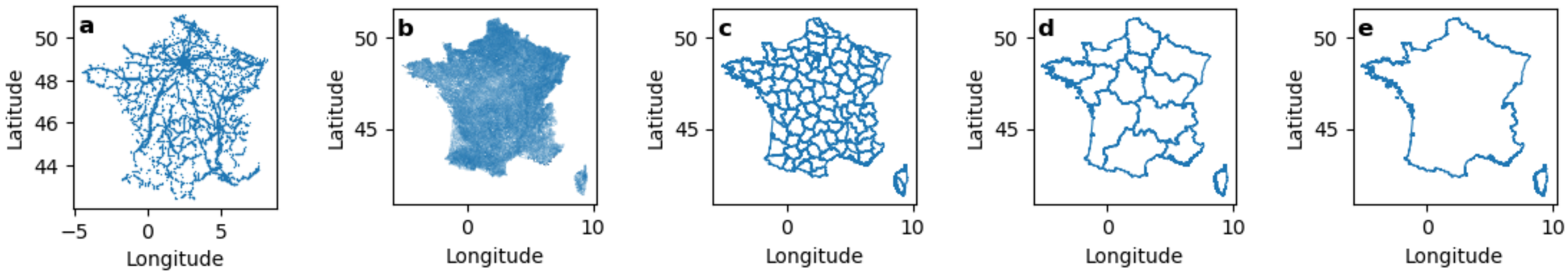

**Fig. 3.4.1.** System Model raw inputs**.** Raw inputs for the System Model at different levels, from an asset, i.e, one point, through a set of assets, to the 'Country' administrative unit. **a.** The SNCF Reseau Road Bridges (SNCF-RRB). One point corresponds to one bridge. **b.** The French 'Communes' administrative units. **c.** The French 'Départements' administrative units. **d.** The French 'Régions' administrative units. **e.** The French 'Country' administrative unit.

### 3.5. Boolean Exposure Model: illustrative case study setup

The Boolean Exposure Model combines two inputs: the hazard outputs from Section 3.3 and the system of interest from Section 3.4. We apply it to the ensemble of SNCF Réseau road bridges, using the nearest SAFRAN grid point to intersect hazard values with physical systems. This instantiation allows us to identify which bridges are exposed to temperature extremes that exceed their design thresholds, directly addressing the problem raised in Section 1. While applicable to all system levels, we focus on the asset level in the presented results.

### 3.6. Boolean Exposure Model: Results

The Boolean Exposure Model reveals a clear trend in Fig.3.6: a significant portion of the SNCF Réseau road bridges across metropolitan France are projected to face extreme temperature conditions that exceed their design thresholds, as defined by Eurocode standards.
The results show that:

- Exposure is widespread and increasing: Under all future climate scenarios (RCP 2.6, 4.5, and 8.5) and across all time periods (2006-2036, 2030-2060, 2054-2084), the portion of exposed bridges increases compared to historical baselines. This indicates a systematic rise in exposure of these road bridges to extreme temperatures due to climate change.
- Spatial non-uniformity is significant: Exposure is not uniform across the territory. Some regions exhibit a higher concentration of exposed assets, suggesting that adaptation strategies may differ geographically.
- Hazard Intensity Thresholds are exceeded differently: The multiclass BEM output (e.g., HIT +2.5°C, etc) highlights that many bridges are not just marginally exposed. Bridges may face substantial exceedances of their design limits. This suggests that some bridges could enter a damage uncertainty zone under future climate conditions.
- By linking exposure to individual bridge locations, the model provides granular information. This information can support prioritization of vulnerability modelling and inspections.

These results reflect a specific modeling configuration that includes a single climate model chain, a Gumbel-based extreme value model, and Eurocode-based thresholds over 30-year periods. While still in the early stages, these findings demonstrate the practical value of the BEM framework. With further refinement and validation, it could enable infrastructure managers to move from climate projection exposure to concrete climate hazards exposure scenarios. These scenarios would be both asset-level and ensemble-level, supporting proactive risk management for physical infrastructures facing climate change hazards.

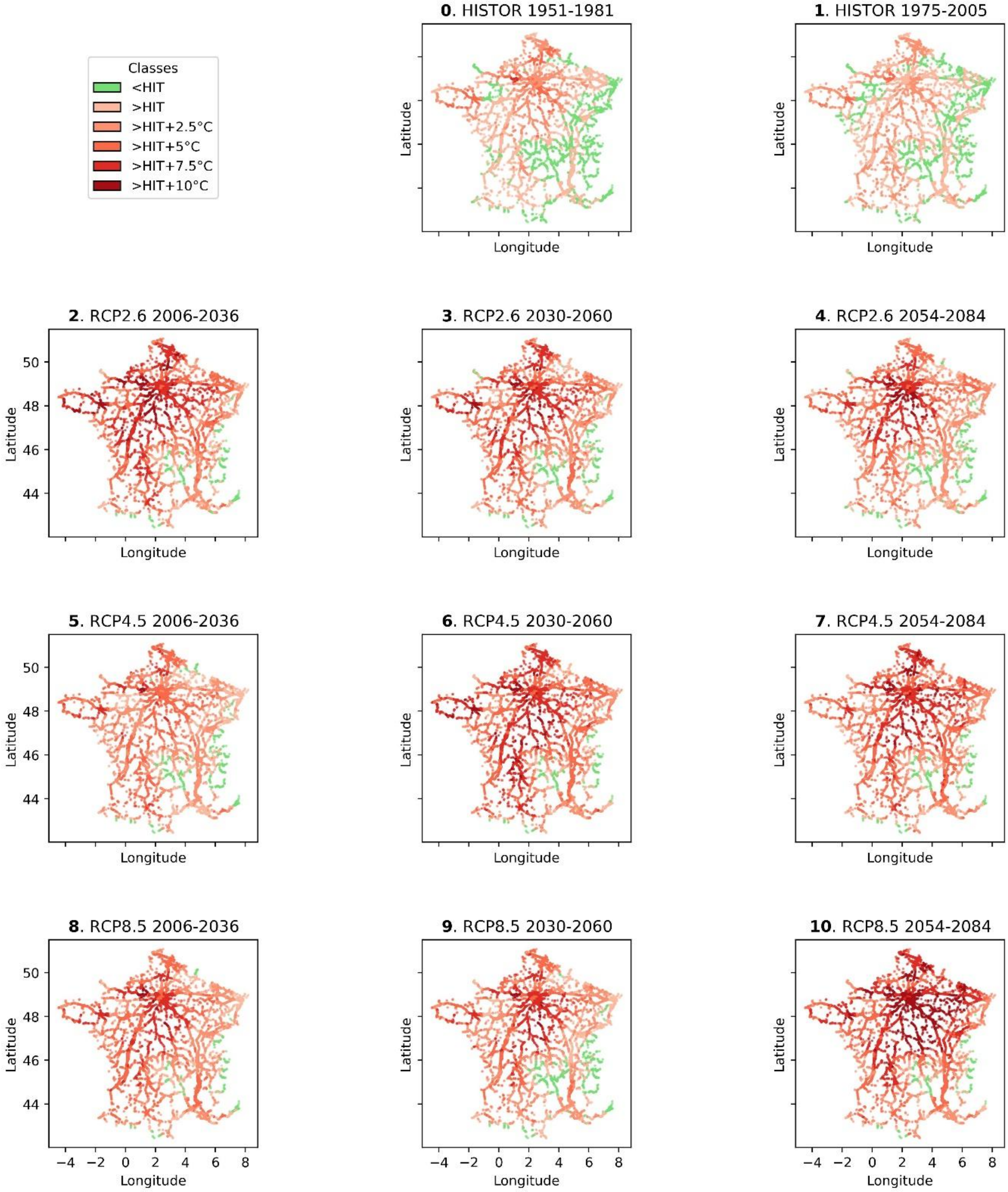

**Fig. 3.6. Boolean Exposure Modeling multiclass output.** These figures represent a system of an ensemble of SNCF Réseau road bridges spread across the entire territory of France, excluding Corsica. The hazard is defined as the exceedance of a 50-year temperature intensity return. This is calculated with a Gumbel distribution based on 30 years of data for each period. Temperature data inputs from CNRM-CM5r1 (GCM), ALADIN6.3v2 (RCM), and ADAMONT FRANCE (BCSD). 'HISTOR' stands for historical.

### 3.7. Exposure Rates Analysis Across System Levels: Results

The Boolean Exposure Model provides outputs that can be shown as an explicit map of exposed assets. The previous results reveal that exposure maps alone may not be sufficient for decision-making. To support infrastructure management and adaptation planning, it is essential to quantify exposure at different system levels.

Aggregated indicators, such as exposure rates, offer a concise way to compare the proportion of exposed assets within larger administrative or organizational units. This section presents two examples of exposure rate analysis. The first example shows the National-Level Exposure Rate. Figure 3.7.1. shows the exposure rate of SNCF Réseau road bridges across the entire French territory, using the Eurocode-based Hazard Intensity Threshold (HIT). The results indicate that:

- In historical periods (1951-1981 and 1975-2005), fewer than 80% of bridges were exposed to extreme temperatures exceeding the HIT.
- In all future periods (2006-2036,2030-2060,2054-2084) and across all RCP scenarios, more than 90% of bridges are projected to be exposed.

These results, reflecting one climate model chain, reveal a clear trend: exposure rates for road bridges increase consistently across future periods and scenarios (more details in Appendix A.2).

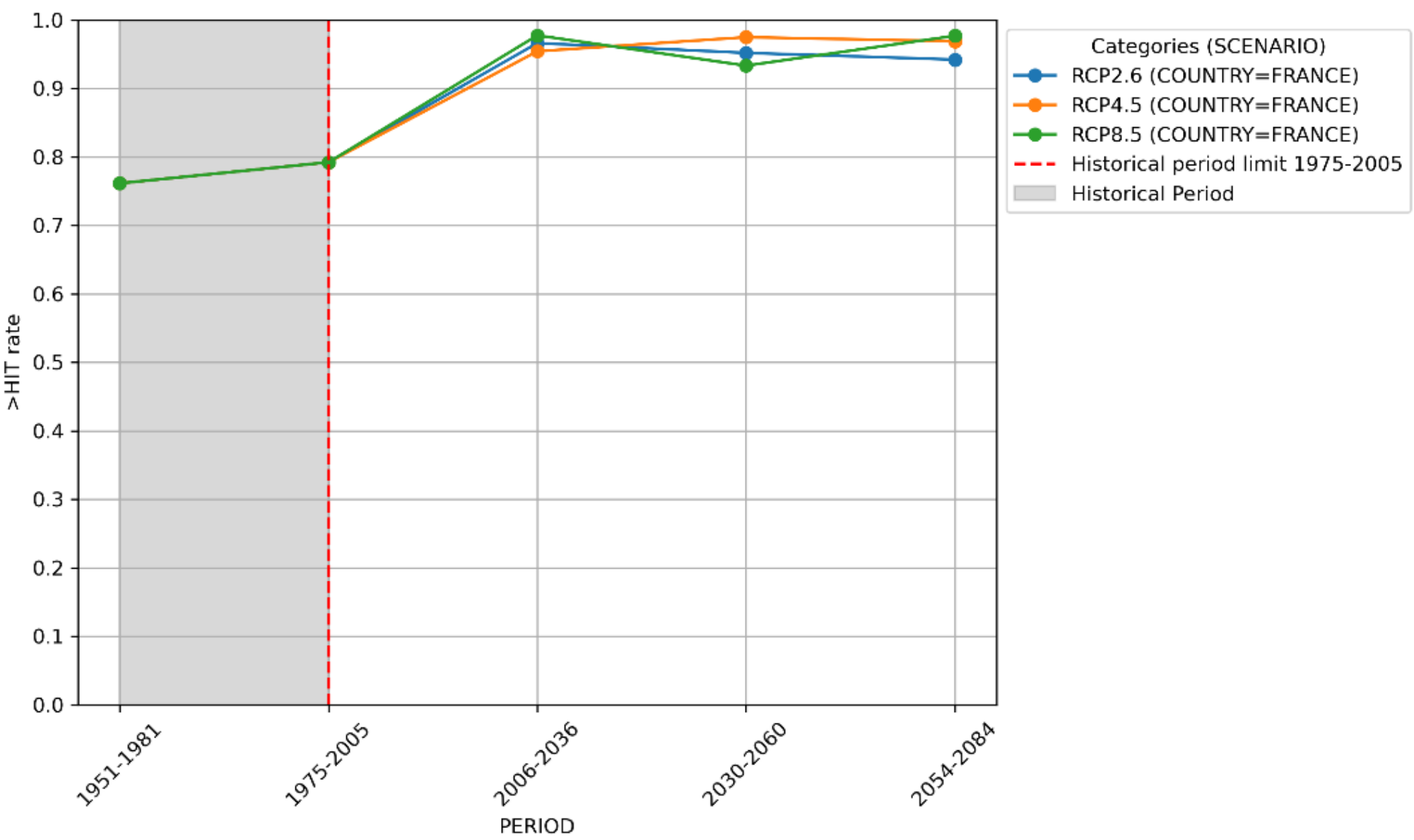

**Fig.3.7.1.** Boolean Exposure Model output rate for a lower level at the bridge level and a high level at the country level. This figure is based on a geographically non-uniform hazard intensity threshold set for the entire territory of France, denoted as 'HIT'. The Boolean Exposure Model is applied here to the system comprising the bridges within the SNCF Réseau Road bridges dataset intersecting the hazard, defined via the BHMP, as the exceedance of the geographically non-uniform HIT by the 50-year return temperature intensity (EVS model: Gumbel). Temperature data inputs from CNRM-CM5r1 (GCM), ALADIN6.3v2 (RCM), and ADAMONT FRANCE (BCSD).

The second example is about the regional-level exposure rate. Figure 3.7.2. presents the same analysis as in the first example, but with exposure rates aggregated at the regional level. This disaggregation enables:

- Regional comparisons: Asset managers can identify which regions have the highest proportion of exposed bridges.
- Targeted actions: Regions with higher exposure rates may require prioritized studies and interventions.
- Verification of regional trend: could help assess whether local patterns align with broader national trends.

This multi-level exposure rate analysis demonstrates the flexibility of the BEM framework in supporting both strategic planning at the national level and operational decision-making at regional or local levels.

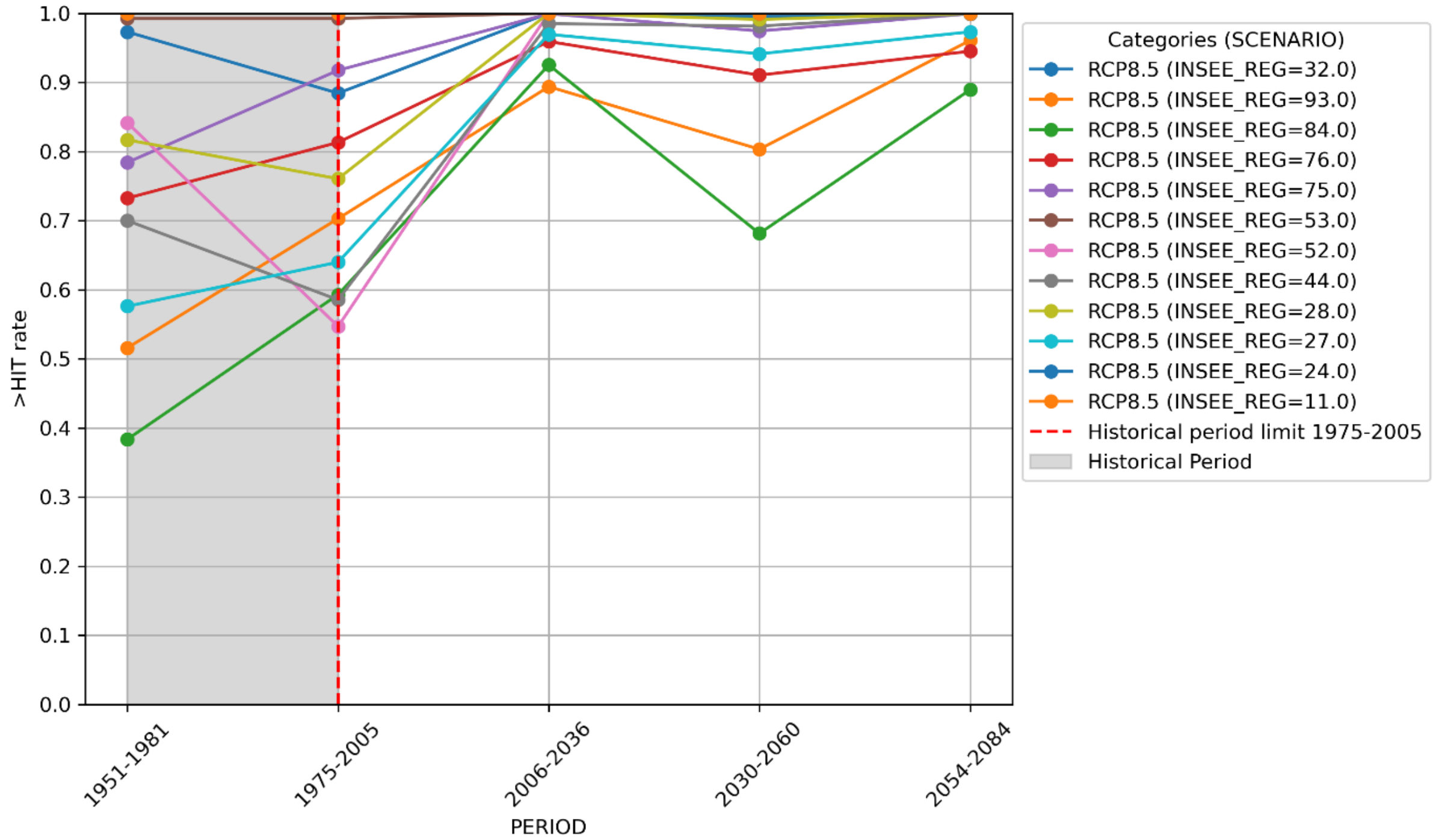

**Fig.3.7.2.** Boolean Exposure Model output rate. The system's lower level is the bridge level; the system's high level is the region level. Each region is represented by an INSEE_REG code; the threshold is 'HIT'. The Boolean Exposure Model is applied to the system comprising the bridges within the SNCF Réseau road bridges dataset, to the extreme temperature hazard. Temperature data inputs from CNRM-CM5r1 (GCM), ALADIN6.3v2 (RCM), and ADAMONT FRANCE (BCSD).

Exposure rates have been presented only for the baseline Hazard Intensity Threshold (HIT). However, the same analytical approach applies to higher thresholds, such as HIT+2.5°C. In practice, asset managers may adopt more conservative thresholds to account for design margins, uncertainty in climate projections, or asset-specific resilience measures. Analyzing exposure rates across multiple thresholds enables prioritization of interventions based on the importance of exceedance, thereby distinguishing between marginal and critical exposure cases. While exposure rates are useful for comparing trends across administrative units, they do not accurately capture the reality of asset management exposure. A region with a high exposure rate but few assets may, at the national level, represent a lower operational risk than a region with a moderate rate but a large number of exposed infrastructures. Therefore, for informed decision-making, exposure rates should be interpreted in conjunction with the absolute number of exposed assets.
Examples of exposed asset counts are provided in Appendix Table A.2. These counts complement the exposure rate analysis, offering a more comprehensive view of the infrastructure systems' exposure to climate hazards.

# 4. Discussion

This work contributes to the transition from raw climate information, including extreme temperature projections, and raw system data, to structured exposure modeling. This paper highlights the importance of identifying the relevant thresholds for defining hazards in the context of physical infrastructure. Our reproducible framework supports this transition, offering a foundation for future decision-making.
The results presented in this study offer a structured and scalable approach to quantifying infrastructure exposure to climate-related hazards. By integrating climate projections, extreme value statistics, and Boolean modeling, the framework enables asset-level and ensemble-level exposure assessments. These assessments currently provide consultative insights. With further refinement, the exposure framework could support actionable decision-making for infrastructure managers, particularly in identifying assets that may exceed design thresholds under future climate scenarios.

The exposure maps and rates reveal consistent trends across scenarios and time periods. They show that a growing proportion of road bridges in metropolitan France are projected to be exposed to extreme temperatures beyond Eurocode-based thresholds. This trend is spatially non-uniform, suggesting that adaptation strategies should be tailored to regional exposure profiles.
The use of multiclass thresholds (e.g., HIT +2.5°C) further enhances the framework's relevance for asset management. Multiclass allows for differentiation between marginal and severe exceedance cases, supporting the prioritization of inspections based on the severity of projected exposure.

However, the results must be interpreted within the scope of the modeling configuration. This study uses a unique climate model chain, a Gumbel-based statistical approach, and Eurocode thresholds applied over 30-year periods. While this setup captures some uncertainty due to the use of multiple scenarios and the inclusion of at least 30 years of data per time period, it does not account for the full range of modeling uncertainty, such as variability across climate models or statistical methods.
The study demonstrates how exposure scenarios can be obtained in a structured manner; however, future work could explore an ensemble of climate models, various infrastructure types, refined thresholds, additional climate change hazards, and an ensemble of extreme modeling approaches.
Importantly, the study does not provide a fixed exposure scenario outcome; it does provide a structured way to obtain exposure scenarios. These scenarios can inform further vulnerability assessment or engineering evaluations.

# 5. Conclusion

This study introduced a structured framework for modelling the exposure of physical infrastructure systems to climate-related hazards. The framework integrates climate projections, extreme value statistics, and Boolean modeling to assess if assets may face hazard intensities that exceed their design thresholds.

Our methodology was applied to a national-scale case study of road bridges in metropolitan France, using Eurocode-based threshold and temperature extremes from the Gumbel distribution. The results demonstrate how exposure scenarios can be quantified across multiple system levels and climate scenarios. The resulting spatially differentiated trends support targeted adaptation strategies.

More broadly, the framework is adaptable to other infrastructure types, hazards, and design standards.
The methodology provides a structured basis for generating exposure scenarios that can be used as pre-analysis for vulnerability assessments and asset management decisions.

Future work should explore various sets of exposure assessments based on multi-model climate ensembles and different types of climate hazards. Additional work should consider alternative or ensemble statistical approaches as well as applications to other physical systems. These directions aim to enhance robustness and support integration into broader reliability and adaptation analyses of physical systems.

# Appendix

**Appendix A.1. Spatial Distribution of the physical system of interest in different administrative unit levels.**

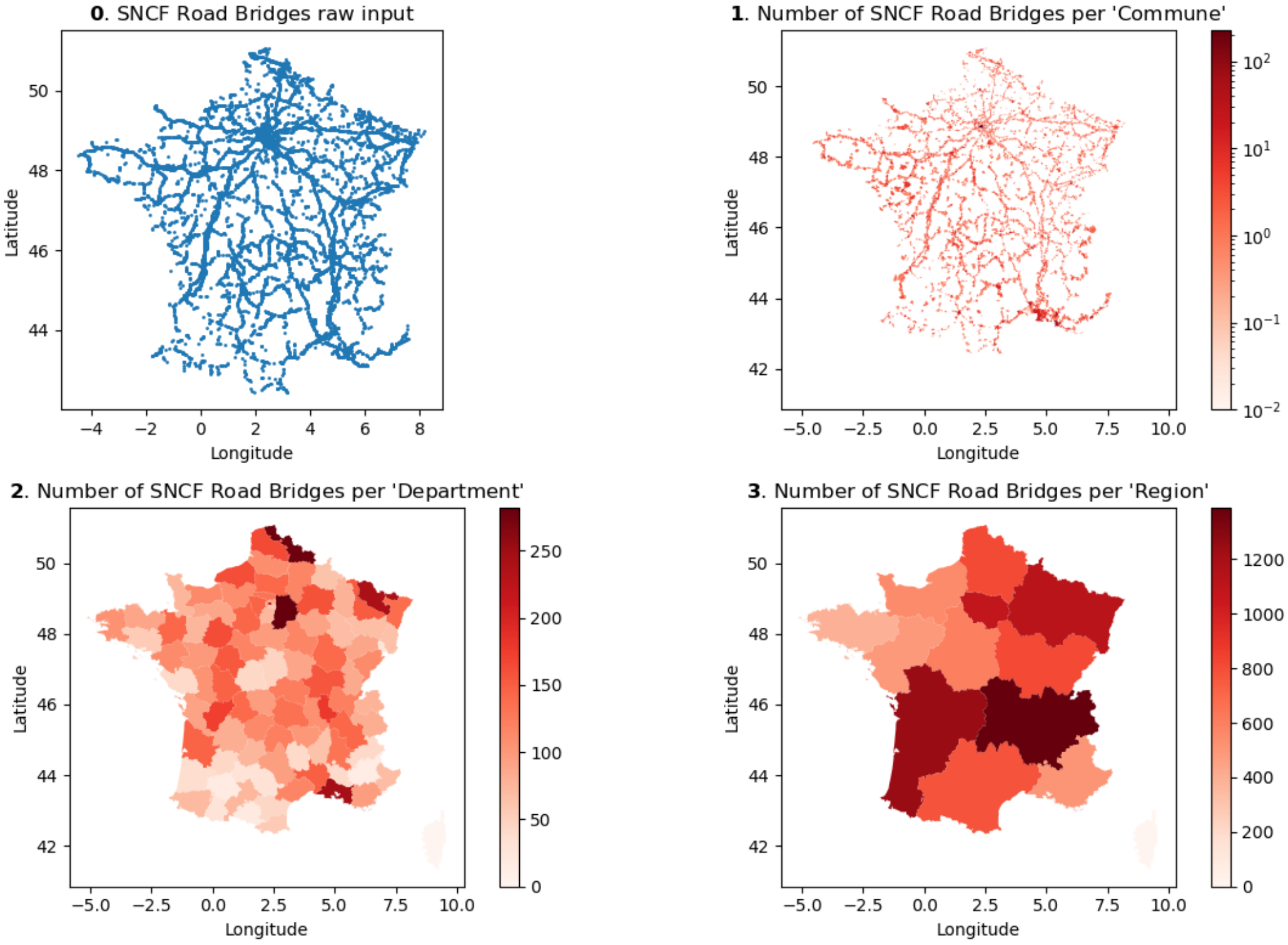

**Fig. A.1.** Spatial distribution of the physical system of interest: SNCF Reseau Road Bridges (SNCF-RRB). **0**. SNCF-RRB raw input with longitude and latitude information alone. **1**. Number of SNCF-RRB per administrative unit: 'Commune'. **2**. Number of SNCF-RRB per administrative unit: 'Département'. **3**. Number of SNCF-RRB per administrative unit: 'Region'.

**Appendix A.2. Aggregated results of the Boolean Exposure Model**

Table A.2 presents the aggregated results of the Boolean Exposure Model applied to the SNCF Réseau Road bridges dataset. For each administrative level (country and selected regions), it reports the number of exposed infrastructures, the total number of infrastructures in the studied sub-ensemble, and the resulting exposure rate. These indicators are computed for multiple time periods under the RCP8.5 scenario. The table highlights both the increasing trend in exposure over time and spatial non-uniformity in the number of exposed assets and exposure rates across regions. For brevity, Table A.2 includes only results for the national level and subset of French regions under a single scenario (RCP8.5). However, the illustrated use case developed in this study enables the same analysis to be conducted across different administrative levels, climate scenarios, and hazard intensity thresholds (e.g., HIT+2.5°C).

| Period / Scenario | | 1951-1981 Historical | 1975-2005 Historical | 2006-2036 RCP8.5 | 2030-2060 RCP8.5 | 2054-2084 RCP8.5 |
|---|---|---|---|---|---|---|
| **LEVEL = COUNTRY** | **type** | | | | | |
| FRANCE | Number of Exposed Systems | 7343 | 7633 | 9580 | 9089 | 9575 |
| | Exposure Rate | 0,75 | 0,78 | 0,98 | 0,93 | 0,98 |
| | Total Number of Systems | | | 9816 | | |
| **LEVEL = REGION** | | | | | | |
| ILE-DE-FRANCE | Number of Exposed Systems | 1090 | 1090 | 1090 | 1090 | 1090 |
| | Exposure Rate | 1 | 1 | 1 | 1 | 1 |
| | Total Number of Systems | | | 1090 | | |
| BOURGOGNE-FRANCHE-COMTE | Number of Exposed Systems | 470 | 522 | 791 | 768 | 794 |
| | Exposure Rate | 0,58 | 0,64 | 0,97 | 0,94 | 0,97 |
| | Total Number of Systems | | | 816 | | |
| AUVERGNE-RHONE-ALPES | Number of Exposed Systems | 533 | 824 | 1286 | 947 | 1236 |
| | Exposure Rate | 0,38 | 0,59 | 0,93 | 0,68 | 0,89 |
| | Total Number of Systems | | | 1390 | | |

**Table. A.2.** Boolean Exposure Model output counts for a lower level at the bridge level and a high level at the country level & at the region level (ILE-DE-FRANCE; BOURGOGNE-FRANCHE-COMTE; AUVERGNE-RHONE-ALPES). The threshold is HIT. The scenario is RCP8.5. The Boolean Exposure Model is applied here to the system comprising the bridges within the SNCF Réseau Road Bridges dataset, to the hazard of return of maximum temperatures (EVS model: Gumbel). Temperature data inputs from CNRM-CM5r1 (GCM), ALADIN6.3v2 (RCM), and ADAMONT FRANCE (BCSD).

# CRediT authorship contribution statement

**Matthieu Dutel**: Methodology & Models, Software, Results & Figures, Writing – original draft, Writing – review & editing. **Didier Soto**: Writing – review & editing, Supervision. **Adam Abdin**: Writing – review & editing, Supervision. **Anne Barros**: Writing – review & editing, Supervision, Project administration.

# Acknowledgement

First author thanks Sixense and its clients, engineers, asset managers, and experts for their advice on HIT choices; thanks also to the LGI team for their research advice.
This paper was written in the context of a CIFRE partnership with CentraleSupélec, Université Paris-Saclay; Resallience by Sixense Engineering, part of the VINCI group; and ANRT.